\documentclass[aps,prl,twocolumn,twoside,superscriptaddress,nofootinbib,preprintnumbers,10pt,floatfix]{revtex4-2}

\usepackage{hyperref}
\usepackage[all]{hypcap}
\usepackage{graphicx}
\usepackage{amsmath,amsfonts,amssymb}
\usepackage{xcolor}
\usepackage{physics}
\usepackage{placeins}

\begin{document}

\title{Koba-Nielsen-Olesen scaling in quark and gluon jets at the LHC}

\author{Xiang-Pan Duan}
\email{xpduan20@fudan.edu.cn}
\affiliation{Key Laboratory of Nuclear Physics and Ion-beam Application~(MOE), Institute of Modern Physics, Fudan University, Shanghai $200433$, China}
\affiliation{Shanghai Research Center for Theoretical Nuclear Physics, NSFC and Fudan University, Shanghai $200438$, China}
\affiliation{Instituto Galego de F\'isica de Altas Enerx\'ias IGFAE, Universidade de Santiago de Compostela, E-15782 Galicia, Spain}

\author{Lin Chen}
\email{lin.chen@usc.es}
\affiliation{Instituto Galego de F\'isica de Altas Enerx\'ias IGFAE, Universidade de Santiago de Compostela, E-15782 Galicia, Spain}

\author{Guo-Liang Ma}
\email{glma@fudan.edu.cn}
\affiliation{Key Laboratory of Nuclear Physics and Ion-beam Application~(MOE), Institute of Modern Physics, Fudan University, Shanghai $200433$, China}
\affiliation{Shanghai Research Center for Theoretical Nuclear Physics, NSFC and Fudan University, Shanghai $200438$, China}

\author{Carlos A. Salgado}
\email{carlos.salgado@usc.es}
\affiliation{Instituto Galego de F\'isica de Altas Enerx\'ias IGFAE, Universidade de Santiago de Compostela, E-15782 Galicia, Spain}

\author{Bin Wu}
\email{b.wu@cern.ch}
\affiliation{Instituto Galego de F\'isica de Altas Enerx\'ias IGFAE, Universidade de Santiago de Compostela, E-15782 Galicia, Spain}

\begin{abstract}
The Koba-Nielsen-Olesen (KNO) scaling of hadron multiplicity distributions, empirically confirmed to hold approximately in $e^+e^-$ collisions and Deep Inelastic Scattering, has been observed to be violated in hadron-hadron collisions. In this work, we show that the universality of KNO scaling can be extended to hadron-hadron collisions when restricted to QCD jets. We present a comprehensive study of KNO scaling in QCD jets produced in proton-proton collisions at the LHC. 
Using perturbative QCD calculations in the double logarithmic approximation and {\tt PYTHIA} simulations, we find that KNO scaling approximately holds for both quark and gluon jets across a broad jet $p_T$ range, from $0.1$ TeV to $2.5$ TeV, at both the parton and hadron levels. Especially, we highlight characteristic differences between the KNO scaling functions of quark and gluon jets, with the quark-jet scaling function lying above that of gluon jets at both low and high multiplicities. This distinction is essential for interpreting inclusive jet data at the LHC. Furthermore, we propose direct experimental tests of KNO scaling in QCD jets at the LHC through quark-gluon discrimination using jet substructure techniques, as demonstrated by applying energy correlation functions to {\tt PYTHIA}-generated data.
\end{abstract}

\maketitle

{\it Introduction.---} Pre-QCD descriptions of hadron production in $e^+e^-$ collisions~\cite{Polyakov:1970lyy} and high-energy hadron collisions~\cite{Koba:1972ng} proposed that hadron multiplicity distributions $P(n)$ exhibit universal scaling as a function of $n/\bar{n}$: $\bar{n} P(n) = \Psi\left({n}/{\bar{n}}\right)$,  
where $\bar{n}$ is the mean multiplicity and $\Psi$ is a universal scaling function. This phenomenon, known as {\it Koba-Nielsen-Olesen (KNO) scaling}, has been a cornerstone of theoretical and experimental investigations in collider physics. KNO scaling has been empirically confirmed to hold approximately in $e^+e^-$ collisions~\cite{TASSO:1983cre, DELPHI:1990ohs, DELPHI:1991qnt}\footnote{
The superposition of jets of different flavors and topologies is expected to violate KNO scaling, as previously discussed in Refs.~\cite{ALEPH:1995qic, Giovannini:1997ce}.
}
and has been reaffirmed in a recent reanalysis of Deep Inelastic Scattering (DIS) data~\cite{H1:2020zpd}. 
However, violations of KNO scaling have been observed in overall charged particle multiplicity distributions in hadron-hadron collisions~\cite{UA5:1985kkp, UA5:1988gup, CMS:2010qvf, Grosse-Oetringhaus:2009eis, ALICE:2017pcy}, raising fundamental questions about its universality in all collision systems.

In perturbative QCD, theoretical studies suggest that KNO scaling in QCD jets should emerge in the asymptotic limit of high virtuality~\cite{Konishi:1979cb, Bassetto:1979nt, Dokshitzer:1982ia, Bassetto:1987fq, Dokshitzer:1991wu, Dokshitzer:1993dc, Malaza:1984vv}, as reviewed in~\cite{Dremin:2000ep}. Hence, a natural observable for testing the universality of KNO scaling in hadron-hadron collisions is hadron multiplicity distributions within QCD jets. These jets are typically initiated by high-virtuality partons, similar to those in $e^+e^-$ collisions and DIS, in contrast to the overall charged hadron multiplicity, which includes contributions from soft processes. The high resolution of LHC detectors, to be further enhanced in the upcoming High-Luminosity LHC era~\cite{Azzi:2019yne}, has led to the renaissance of jet physics~\cite{Marzani:2019hun, Larkoski:2024uoc}, providing a unique opportunity to investigate KNO scaling in jets with unprecedented accuracy.

Despite extensive studies of jet physics at the LHC using jet substructure techniques~\cite{Marzani:2019hun, Larkoski:2024uoc}, KNO scaling within QCD jets remains largely untested. Recently, {\tt PYTHIA} simulations suggest that KNO scaling is approximately preserved within jets, with deviations attributed to multi-parton interactions (MPI) and hadronization effects~\cite{Vertesi:2020utz}. Furthermore, an analysis of KNO scaling in LHC jet measurements~\cite{ATLAS:2011eid, ATLAS:2019rqw} has been presented in Ref.~\cite{Germano:2024ier}. These developments raise a critical question: does KNO scaling in QCD jets indeed manifest in existing LHC data, and how can it be further tested at the LHC? Answering this requires an in-depth investigation of KNO scaling in the modern era of jet physics, combining theoretical studies with high-precision experimental measurements. 

In this work, we present an up-to-date investigation of KNO scaling in QCD jets at LHC energies, advancing previous studies on three key fronts: (i) the first direct verification of KNO scaling for both quark and gluon jets in the double logarithmic approximation (DLA), incorporating color coherence effects~\cite{Dokshitzer:1987nm, Dokshitzer:1991wu}, through direct computation of parton multiplicity distributions up to $n = 1000$ across LHC-relevant $p_T$ ranges; (ii) an alternative interpretation of the observed features in inclusive charged particle multiplicity distributions within leading dijets at the LHC~\cite{ATLAS:2019rqw}, compared to that in Ref.~\cite{Germano:2024ier}, based on distinct KNO scaling functions for quark and gluon jets observed in {\tt PYTHIA} simulations; and (iii) the first concrete demonstration—using {\tt PYTHIA}—of the feasibility of directly probing KNO scaling in quark and gluon jets at the LHC using jet substructure techniques, exemplified through energy correlation functions~\cite{Larkoski:2013eya}.

{\it KNO scaling within QCD jets: a general argument.---} 
In the asymptotic limit of high virtuality, KNO scaling in multiplicity distributions within QCD jets naturally follows the definition of {\it generating functions} (GFs)~\cite{Konishi:1979cb, Bassetto:1979nt, Dokshitzer:1982ia, Dokshitzer:1991wu, Dremin:2000ep}:
\begin{align}
\label{eq:GF_def}
    Z_a(u, Q) \equiv \sum_{n=0}^{\infty} u^n P_a(n, Q), 
\end{align}
where the subscript $a$ labels the parton that initiates the jet, and the initial jet scale is given by $Q = p_T R$, with $R$ the jet radius and $p_T$ the jet transverse momentum. Here, the multiplicity distribution $P_a(n, Q)$ represents the probability of finding $n$ particles in a jet initiated by parton $a$. In the asymptotic limit $Q \to \infty$, Eq.~\eqref{eq:GF_def} implies~\cite{Dokshitzer:1982ia, Dokshitzer:1991wu}
\begin{align}\label{eq:kno2}
    \Phi_a(\beta) \equiv& \lim\limits_{Q\to \infty} Z_a\left(e^{-\frac{\beta}{\bar{n}_a}}, Q\right)
    =\lim\limits_{Q\to \infty} \sum_{n=0}^{\infty}  \frac{e^{-\beta \frac{n}{\bar{n}_a}}}{\bar{n}_a}\notag\\
    &\times [\bar{n}_a P_a(n, Q)]
    = \int_0^{\infty} \dd x \, \Psi_a(x) e^{-\beta x},
\end{align} 
where $x \equiv n/\bar{n}_a$, and the sum over $n$ has been converted into an integral over $x$, given that $\bar{n}_a$ grows with $Q$ in QCD. 
Accordingly, KNO scaling is expected to hold in this asymptotic limit~\cite{Konishi:1979cb, Dokshitzer:1982ia, Dokshitzer:1991wu, Dokshitzer:1993dc}, with the scaling functions $\Psi_a(x)$ obtained as the inverse Laplace transform of $\Phi_a(\beta)$.

{\it Asymptotic KNO scaling within QCD jets in DLA.---} The asymptotic KNO scaling functions in QCD have not been fully evaluated beyond DLA. Even within DLA, their explicit forms in parton multiplicity distributions are only approximately known, with results for quark jets provided in~\cite{Bassetto:1987fq} and gluon jets in~\cite{Dokshitzer:1993dc}. Here, we elucidate the differences in the shapes of the asymptotic scaling functions for quark and gluon jets in the limits of small and large $x = n/\bar{n}$.

The GFs for parton multiplicity distributions in DLA reduce to~\cite{Dokshitzer:1982ia, Dokshitzer:1991wu}: 
\begin{align}
    Z_a(u, y) = u \exp \big\{ c_a \int\limits_0^y \dd\bar{y} \, (y - \bar{y}) \gamma^2_0 [ Z_g(u,\bar{y}) - 1 ] \big\},
\end{align} 
where $y \equiv \ln(Q/Q_0)$, $\bar{y} \equiv \ln(k_\perp/Q_0)$, with $Q_0$ standing for the transverse momentum of final-state partons with respect to the jet direction. 
Here, $\gamma_0^2$ depends on $\bar{y}$ through the running coupling: $\gamma_0^2 = {2N_c \alpha_s(k_\perp^{2})}/{\pi}$, and $c_a = c_g \equiv {C_A}/{N_c} = 1$ for gluon jets and $c_a = c_q \equiv {C_F}/{N_c} = {4}/{9}$ for quark jets, respectively. 

From Eq.~\eqref{eq:kno2}, one has $\Phi_q(\beta) = \Phi_g^{c_q}(\beta / c_q)$. Accordingly, by performing the inverse Laplace transform of $\Phi_q(\beta)$ for small and large $|\beta|$, respectively, and using the known asymptotic behavior of $\Phi_g(\beta)$~\cite{Dokshitzer:1982ia, Dokshitzer:1991wu}, we obtain the following results: For $x \gg 1$,
\begin{align}
\label{eq:KNO_DLA_largex}
    \Psi_q(x) &\approx 0.614 \beta_0^{\frac{8}{9}} \frac{e^{-\frac{C_F}{N_c} \beta_0 x}}{x^\frac{1}{9}}~\textbf{vs}~ \Psi_g(x) \approx 2\beta_0^2 x e^{-\beta_0 x}, 
\end{align} 
where the integration constant $\beta_0 \approx 2.553$. For $x \ll 1$, 
\begin{align}
\label{eq:KNO_DLA_smallx}
    \Psi_q(x) &\sim \frac{1}{x} e^{-\frac{1}{2}\frac{C_F}{N_c} \ln^2 x}~~\textbf{vs}~~ \Psi_g(x) \sim \frac{1}{x} e^{-\frac{1}{2} \ln^2 x}. 
\end{align} 
These results reveal that the quark-jet KNO function lies above that of gluon jets in both the low and high $n/\bar{n}$ regions. Additionally, the quark-jet scaling function peaks lower and is shifted to the left relative to that of gluon jets, ensuring probability conservation. This qualitative behavior persists beyond DLA in hadron multiplicity distributions, as we confirm below.

{\it Verifying KNO scaling within QCD jets in DLA.---}
To directly test the validity of the asymptotic KNO scaling functions at LHC energies, we numerically solve $P_a(n)$ using the following iterative relation: 
\begin{align}
    P_a(1) &= \exp \left\{ -c_a \int_0^y \dd\bar{y} \, (y - \bar{y}) \gamma^2_0 \right\}, \notag \\
    P_a(n+1) &= \sum\limits_{k=1}^n \frac{k}{n} P_a(n+1-k) \notag \\
    &\quad \times c_a \int_0^y \dd\bar{y} \, (y - \bar{y}) \gamma^2_0 P_g(k, \bar{y}),
\end{align} 
along with the mean parton multiplicities $\bar{n}_a = \sum_n n P_a(n)$, which satisfy $\bar{n}_q = c_q (\bar{n}_g - 1) + 1$.

\begin{figure}[!htbp]
    \includegraphics[height=0.22\textheight]{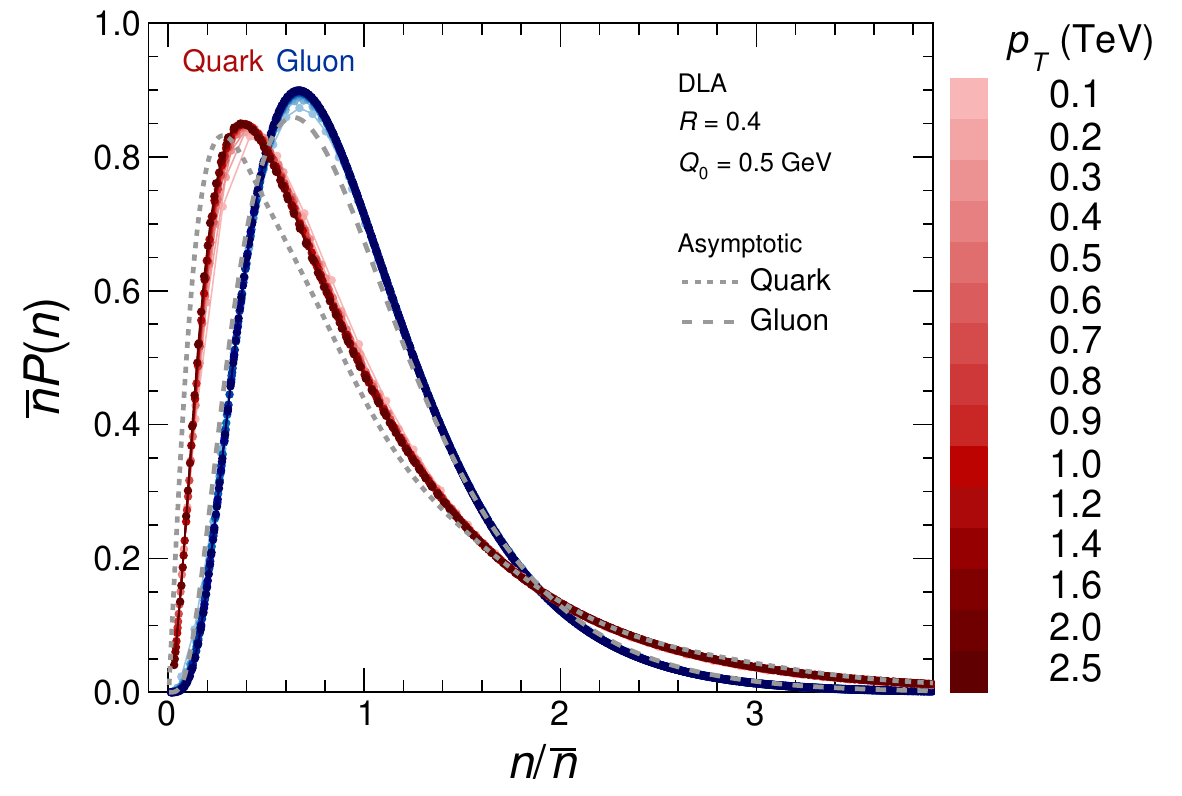}
    \caption{KNO scaling in parton multiplicity distributions within DLA. Over a broad jet $p_T$ range of $p_T = 0.1$–$2.5$ TeV, $\bar{n} P(n)$ for both quark and gluon jets (computed up to $n = 1000$) converge to well-defined scaling functions, with minor deviations observed at lower $p_T$. They deviate slightly from the asymptotic scaling functions (dashed) approximated in Refs.~\cite{Bassetto:1987fq, Dokshitzer:1993dc}. Here, $R = 0.4$ and $Q_0 = 0.5$ GeV.
    }
    \label{fig:KNO_DLA}
\end{figure}

This iterative approach enables a first-principles numerical test of KNO scaling in DLA. In Fig.~\ref{fig:KNO_DLA}, we present $\bar{n}P(n)$ as a function of $n/\bar{n}$ for both quark and gluon jets, computed up to $n = 1000$, with $Q_0 = 0.5$ GeV and $R = 0.4$. These results are compared to the asymptotic KNO scaling functions approximated in Refs.~\cite{Bassetto:1987fq, Dokshitzer:1993dc}, which are matched by the analytic forms in Eqs.~\eqref{eq:KNO_DLA_largex} and \eqref{eq:KNO_DLA_smallx} at large and small $n/\bar{n}$, respectively. Over the jet $p_T$ range of 0.1--2.5 TeV, we find that $\bar{n} P(n)$ for both quark and gluon jets converge to well-defined scaling functions, demonstrating the emergence of KNO scaling across a wide range of jet $p_T$ relevant to LHC energies. Minor discrepancies are observed only at lower $p_T$. These functions exhibit slight deviations from the approximate asymptotic KNO scaling functions in~\cite{Bassetto:1987fq, Dokshitzer:1993dc}. Our results remain robust against variations in the infrared cutoff $Q_0$, provided $Q_0 \sim 1$ GeV, while convergence improves if $Q_0$ becomes smaller, indicating their infrared safety.

{\it KNO scaling in inclusive multiplicity distributions within QCD jets.---} Inclusive charged particle multiplicities within jets have been measured at the LHC~\cite{ATLAS:2011eid, ATLAS:2019rqw}. Assuming KNO scaling for both quark and gluon jets, the inclusive multiplicity distributions for all the selected jets can be expressed as
\begin{align}\label{eq:Pall}
    P(n) &= r_q P_q(n) + r_g P_g(n) = r_q\frac{\Psi_q(x)}{\bar{n}_q} + r_g\frac{\Psi_g(x)}{\bar{n}_g},
\end{align}
where $r_a \equiv \frac{\dd \sigma_a}{\dd p_T}/(\frac{\dd \sigma_q}{\dd p_T }+ \frac{\dd \sigma_g}{\dd p_T})$ represents the fraction of jets initiated by parton $a$ among all the selected jets produced at given jet $p_T$, and $\dd\sigma_a/\dd p_T$ denotes the corresponding jet cross section initiated by parton $a$. 

As a concrete example, we examine inclusive multiplicity distributions in leading dijets for jet transverse momenta in the range $0.1$–$2.5$ TeV in $pp$ collisions at $\sqrt{s} = 13$ TeV, as studied in Ref.~\cite{ATLAS:2019rqw}. To establish a baseline, we start with the leading order (LO) calculation for inclusive parton multiplicity distributions in perturbative QCD, where the cross section is factorized into LO partonic dijet cross sections and the DLA results for $P_a(n)$, as shown in Fig.~\ref{fig:KNO_DLA}. 

\begin{figure}[!htbp]
    \includegraphics[height=0.22\textheight]{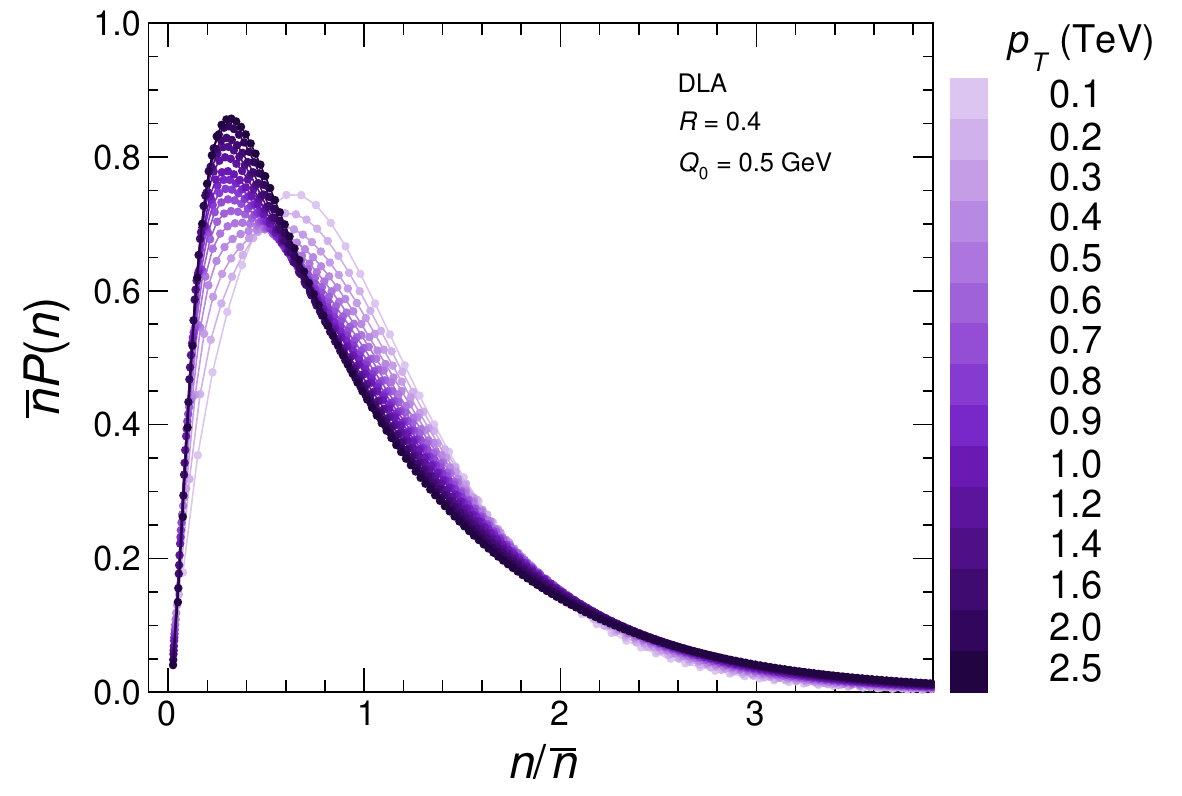}
    \caption{Inclusive parton multiplicity distributions for dijets in DLA. Here, $\bar{n}$ and $P(n)$ represent the averages over leading dijets produced at a given $p_T$ within rapidity range $|\eta|<2.1$ in proton-proton collisions at $\sqrt{s} = 13$ TeV. These distributions are calculated using the mean multiplicities and multiplicity distributions for quark and gluon jets, as shown in Fig.~\ref{fig:KNO_DLA}.
    }
    \label{fig:KNO_DLAAll}
\end{figure}

Fig.~\ref{fig:KNO_DLAAll} shows the rescaled inclusive parton multiplicity distributions $\bar{n} P(n)$ as a function of $n/\bar{n}$, where $\bar{n} = r_q \bar{n}_q + r_g \bar{n}_g$. Here, $\bar{n}$ and $P(n)$ represent the averages over dijets produced at a given $p_T$ within rapidity range $|\eta|<2.1$. The parton distribution functions used in these calculations are taken from CT18NLO via {\tt LHAPDF}~\cite{Buckley:2014ana}, evaluated at the scale given by jet $p_T$. Notably, the distributions for different $p_T$ values do not converge to a single scaling function, highlighting the impact of distinct KNO scaling behaviors in quark and gluon jets on inclusive multiplicity distributions. 

The quantitative features shown in Fig.~\ref{fig:KNO_DLAAll} can be understood directly from Eq.~\eqref{eq:Pall}. This behavior arises because gluon jets dominate at lower $p_T$, while quark jets become increasingly significant at higher $p_T$. Specifically, we find that $\bar{n}P(n) = \Psi(x) \approx 0.4\Psi_q(x_q) + 0.7\Psi_g(x_g)$ at $p_T = 0.1$ TeV, with $r_g \approx 0.78$.
While $\bar{n}P(n) = \Psi(x) \approx 1.0\Psi_q(x_q) + 0.1\Psi_g(x_g)$ at $p_T = 2.5$ TeV, with $r_q \approx 0.81$. Here, $x_a \equiv n/\bar{n}_a$. Thus, $\bar{n}P(n)$ is primarily shaped by quark jets at high $p_T$ and by gluon jets at low $p_T$.

While recent DLA calculations provide a consistent explanation of observed KNO scaling in DIS~\cite{Liu:2022bru, Liu:2023eve}, we find that they are insufficient to describe inclusive charged particle multiplicity distributions at the LHC~\cite{ATLAS:2011eid, ATLAS:2019rqw}. In the absence of complete results for $P_a(n)$ beyond DLA in QCD, we instead use {\tt PYTHIA}~\cite{Bierlich:2022pfr} simulations to extend our analysis to experimental data. Our event selection follows Ref.~\cite{ATLAS:2019rqw}, with the two leading jets reconstructed using the anti-$k_t$ algorithm~\cite{Cacciari:2008gp} ($R = 0.4$) in the {\tt FastJet} package~\cite{Cacciari:2011ma}, requiring $|\eta| < 2.1$ and their momentum ratio ${p_T^\text{lead}}/{p_T^{\text{sublead}}} < 1.5$.

\begin{figure}[!htbp]
    \centering
    \includegraphics[height=0.33\textheight]{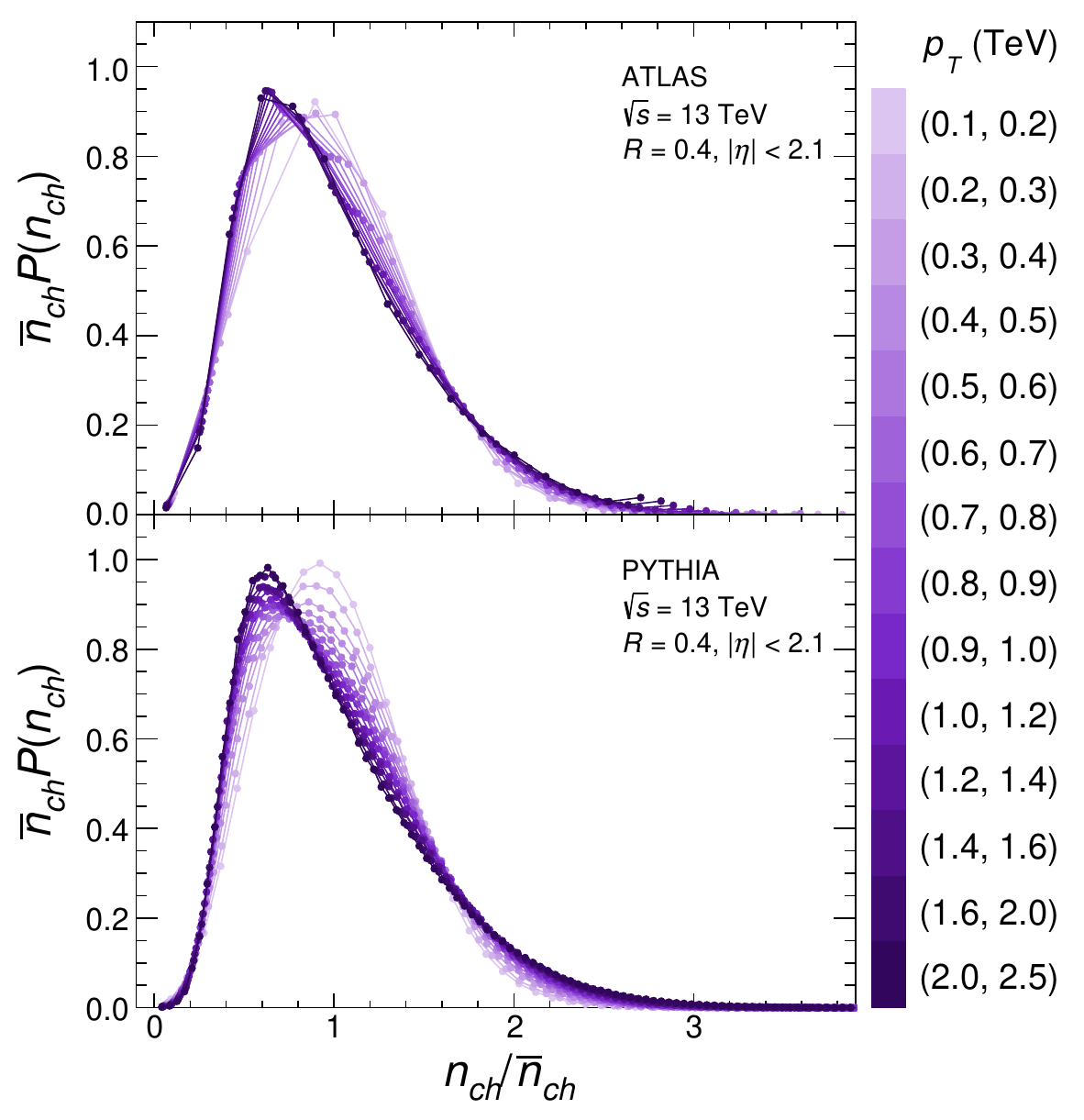}
    \caption{Inclusive charged particle multiplicity distributions in leading dijets at the LHC. The top panel presents experimental data for $\bar{n}_{ch}P(n_{ch})$ as a function of $n_{ch}/\bar{n}_{ch}$ from Ref.~\cite{ATLAS:2019rqw}, with a bin size of $\Delta n_{ch} = 4$, where uncertainty bands are omitted for clarity. The bottom panel displays our \texttt{PYTHIA} results for inclusive charged hadron multiplicity distributions, with results shown for each individual value of $n_{ch}$.}
    \label{fig:KNO_Ch}
\end{figure}

\begin{figure*}[htbp]
    \centering
    \includegraphics[height=0.22\textheight]{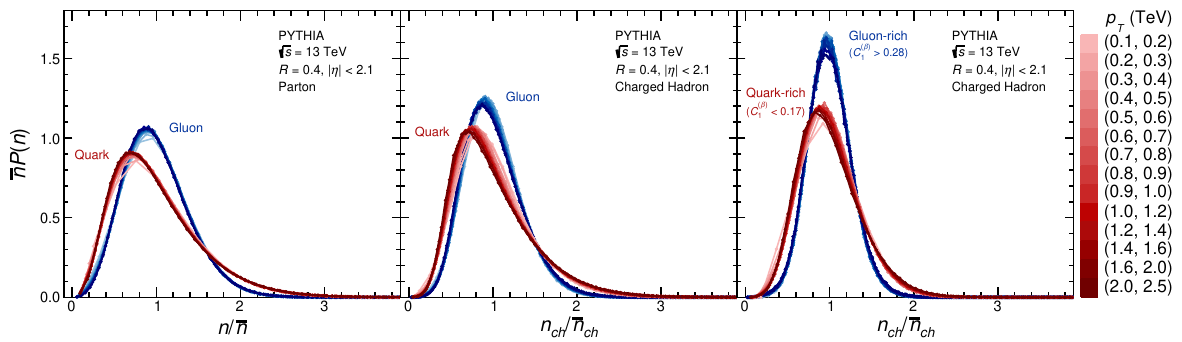}
    \caption{KNO scaling within quark and gluon jets using {\tt PYTHIA}. The left panel displays parton-level results for quark and gluon jets without MPI effects. The central panel shows charged hadron results incorporating MPI effects, with noticeable $p_T$ dependence arising mostly from hadronization. The right panel presents charged hadron results for quark-rich and gluon-rich jets, discriminated using the energy correlation functions~\cite{Larkoski:2013eya}.}
    \label{fig:KNO_MPI_Hadronization_ECFs}
\end{figure*}

The top panel of Fig.~\ref{fig:KNO_Ch} presents the measured inclusive charged particle multiplicity distributions $\bar{n}_{ch}P(n_{ch})$ versus $n_{ch}/\bar{n}_{ch}$ from Ref.~\cite{ATLAS:2019rqw} (with a bin size of $\Delta n_{ch} = 4$). This plot shows that the central values of the experimental data exhibit an evolution from low to high $p_T$ qualitatively similar to our DLA results shown in Fig.~\ref{fig:KNO_DLAAll}, although this trend is partially obscured by experimental uncertainties (not displayed in this plot). Such qualitative similarity is confirmed by our {\tt PYTHIA} results for inclusive charged hadron multiplicity distributions (bottom panel). Notably, the difference between low- and high-$p_T$ regimes is even more pronounced than that shown in the top panel, with the peak position shifting from $n_{ch}/\bar{n}_{ch} \approx 0.92$ at the lowest $p_T$ bin to $n_{ch}/\bar{n}_{ch} \approx 0.63$ at the highest $p_T$ bin.

The key factor in explaining the above observation is the distinction between quark and gluon jets, as evident in Eq.~\eqref{eq:Pall}. To justify this interpretation, we need first to examine KNO scaling in charged hadron multiplicity distributions separately for quark and gluon jets using {\tt PYTHIA}. Jets are discriminated based on their proximity to an initiating parton produced in a $2\rightarrow2$ hard scattering process in {\tt PYTHIA}. We have verified that our method for discriminating quark and gluon jets yields results consistent with those obtained using the methods in Refs.~\cite{Larkoski:2013eya, ATLAS:2019rqw}.

KNO scaling holds approximately for both quark and gluon jets in charged hadron multiplicity distributions simulated with {\tt PYTHIA}. As shown in Fig.~\ref{fig:KNO_MPI_Hadronization_ECFs} (left panel), parton multiplicity distributions in {\tt PYTHIA} exhibit KNO scaling to a degree comparable with our DLA results. The central panel reveals that the rescaled distributions $\bar{n}_{ch}P(n_{ch})$ for both quark and gluon jets display a modest yet more discernible $p_T$ dependence compared to their parton-level counterparts. It primarily arises from hadronization, with MPI playing a secondary role (see also~\cite{Vertesi:2020utz}). Since this $p_T$ dependence does not obscure the distinction between quark and gluon jets, it hence justifies our interpretation that the observed evolution of the peak of $\bar{n}_{ch}P(n_{ch})$ from low $p_T$ to high $p_T$ reflects the transition from gluon-jet-dominated to quark-jet-dominated KNO scaling behavior.

Our interpretation—corroborated by consistent trends at both parton and hadron levels—provides an alternative perspective on KNO scaling, revealing features implicitly encoded in existing LHC data that have not previously been identified in Ref.~\cite{Germano:2024ier}, which did not distinguish between quark and gluon jet contributions and thus overlooked key structural aspects of the scaling behavior.

{\it KNO scaling with quark-gluon discrimination.---} KNO scaling within QCD jets at the LHC can be tested experimentally using jet substructure techniques. A wide range of quark–gluon discriminants is available, including the topic modeling method~\cite{Metodiev:2018ftz, Komiske:2018vkc} adopted by ATLAS~\cite{ATLAS:2019rqw}, as recently reviewed in Refs.~\cite{Marzani:2019hun, Larkoski:2024uoc} and references therein. Although current measurements are limited by experimental uncertainties and coarse binning in $n_{ch}$~\cite{ATLAS:2019rqw}, these methods, capable of isolating quark and gluon jets without imposing additional constraints on phase space, offer a promising avenue to test experimentally whether the KNO scaling observed in {\tt PYTHIA} simulations—illustrated in the central panel of Fig.~\ref{fig:KNO_MPI_Hadronization_ECFs}—is realized in actual data.

As a complementary approach, we propose exploring and testing KNO scaling by employing energy correlation functions (ECFs)\cite{Larkoski:2013eya} to discriminate between quark and gluon jets, as demonstrated using {\tt PYTHIA}-generated data below. These jets are selected within a constrained phase space, for which the presence of KNO scaling has not been established in the existing literature.

For this analysis, we utilize the energy correlation double ratio $C_{1}^{(\beta)} = {{\rm ECF}(2,\beta)}/{{\rm ECF}(1,\beta)^{2}}$ with $\beta = 0.2$~\cite{Larkoski:2013eya}. 
Here, the ECFs are defined in terms of $p_{Ti}$, the transverse momentum of jet constituent $i$, and $R_{ij}$, the distance between jet constituents $i$ and $j$ in the rapidity-azimuth angle plane: ${\rm ECF}(1,\beta) = \sum_{i } p_{Ti}$ and ${\rm ECF}(2,\beta) = \sum_{i<j} p_{Ti} p_{Tj} (R_{ij})^{\beta}$. We investigate two classes of jets: 1) {\it quark-rich jets}, defined as those with $C_1^{(\beta)} < 0.17$; and 2) {\it gluon-rich jets}, defined as those with $C_1^{(\beta)} > 0.28$. By checking the partonic origin of each jet in this classification within {\tt PYTHIA}, we find that the fraction of quark jets among all quark-rich jets increases from 76\% at the lowest $p_T$ bin to 94\% at the highest $p_T$ bin, while the fraction of gluon jets among all gluon-rich jets decreases from 88\% to 52\%. For $p_T \leq 0.7$~TeV, the fraction of gluon jets among all gluon-rich jets remains above 75\% across all $p_T$ bins.

The right panel of Fig.~\ref{fig:KNO_MPI_Hadronization_ECFs} shows that KNO scaling holds for quark-rich and gluon-rich jets in our {\tt PYTHIA}-generated data to an extent comparable to that for quark and gluon jets shown in the central panel. Although the KNO scaling functions for quark-rich and gluon-rich jets differ from those in the two left panels of Fig.~\ref{fig:KNO_MPI_Hadronization_ECFs} and from our DLA results, they exhibit similar qualitative characteristics, with the KNO scaling function of quark-rich jets lying above that for gluon-rich jets at both small and large $n_{ch}/\bar{n}_{ch}$. 

{\it Conclusions and outlook.---}Our study, combining double logarithmic approximation (DLA) calculations in perturbative QCD with \texttt{PYTHIA} simulations, shows that both quark and gluon jets exhibit KNO scaling in multiplicity distributions at both parton and hadron levels. Crucially, these jets retain distinct scaling functions at LHC energies, which are essential for understanding inclusive charged particle multiplicity distributions in leading dijets~\cite{ATLAS:2011eid, ATLAS:2019rqw}. Furthermore, we have demonstrated--for the first time--that KNO scaling in QCD jets can be directly probed using jet substructure techniques, exemplified through energy correlation functions~\cite{Larkoski:2013eya}. Our findings urge immediate experimental scrutiny of KNO scaling directly at the LHC.

Moreover, this work motivates several avenues for future theoretical studies. First, the distinct KNO scaling functions for quark and gluon jets in $e^+e^-$ collisions and DIS~\cite{Liu:2022bru, Liu:2023eve} merit further exploration. Second, while our {\tt PYTHIA} simulations show that the qualitative differences between quark and gluon scaling functions persist beyond DLA and exhibit limited sensitivity to non-perturbative effects, a complete perturbative QCD calculation beyond DLA remains absent. Such a calculation is expected to be crucial for resolving discrepancies between DLA predictions and experimental data~\cite{Malaza:1984vv, Dokshitzer:1993dc, DCMSW2025}. Third, although the DLA results demonstrate the infrared safety of KNO scaling functions despite the infrared sensitivity of parton multiplicities, an examination of their infrared safety at all orders is required. This also raises important questions regarding hadronization effects~\cite{Azimov:1984np, Kang:2023zdx}. Approaches using infrared-safe multiplicity definitions~\cite{Medves:2022ccw, Medves:2022uii} could provide additional insights. Finally, the implications of our findings for heavy-ion collisions~\cite{DiasdeDeus:1997di, Ma:2003vs, Ma:2004eqa, Armesto:2008qe, Armesto:2009aqm, Escobedo:2016vba} are intriguing, and it is worth investigating whether the universality of KNO scaling extends to high-energy nuclear collisions.

{\it Acknowledgments.---}
We thank N\'estor Armesto for helpful discussions. 
This work is supported by the European Research Council project ERC-2018-ADG-835105 YoctoLHC, the project Mar\'{\i}a de Maeztu CEX2023-001318-M financed by MCIN/AEI/\-10.13039/\-501100011033, by European Union ERDF, and by the Spanish Research State Agency under project PID2020-119632GBI00 and PID2023-152762NB-I00.
X.D.~is supported by the China Scholarship Council under Grant No. 202306100165. 
L.C.~is supported by the Marie Sk\l{}odowska-Curie Actions Postdoctoral Fellowships under Grant No.~101210595. 
X.D. and G.M.~are supported by the National Natural Science Foundation of China under Grants No.12147101, No. 12325507, the National Key Research and Development Program of China under Grant No. 2022YFA1604900, and the Guangdong Major Project of Basic and Applied Basic Research under Grant No. 2020B0301030008. 
B.W. acknowledges the support of the Ram\'{o}n y Cajal program with the Grant No. RYC2021-032271-I and the support of Xunta de Galicia under the ED431F 2023/10 project.

\bibliographystyle{apsrev4-2}
\bibliography{ref.bib}

@article{Polyakov:1970lyy,
    author = "Polyakov, A. M.",
    title = "{A Similarity hypothesis in the strong interactions. 1. Multiple hadron production in e+ e- annihilation}",
    journal = "Zh. Eksp. Teor. Fiz.",
    volume = "59",
    pages = "542--552",
    year = "1970",
    url = "http://jetp.ras.ru/cgi-bin/e/index/e/32/2/p296?a=list",
}

@article{Koba:1972ng,
    author = "Koba, Z. and Nielsen, Holger Bech and Olesen, P.",
    title = "{Scaling of multiplicity distributions in high-energy hadron collisions}",
    doi = "10.1016/0550-3213(72)90551-2",
    journal = "Nucl. Phys. B",
    volume = "40",
    pages = "317--334",
    year = "1972"
}

@article{TASSO:1983cre,
    author = "Althoff, M. and others",
    collaboration = "TASSO",
    title = "{Jet Production and Fragmentation in e+ e- Annihilation at 12-GeV to 43-GeV}",
    reportNumber = "DESY-83-130",
    doi = "10.1007/BF01547419",
    journal = "Z. Phys. C",
    volume = "22",
    pages = "307--340",
    year = "1984"
}

@article{DELPHI:1990ohs,
    author = "Abreu, P. and others",
    collaboration = "DELPHI",
    title = "{Charged particle multiplicity distributions in Z0 hadronic decays}",
    reportNumber = "CERN-PPE-90-173",
    doi = "10.1007/BF01474073",
    journal = "Z. Phys. C",
    volume = "50",
    pages = "185--194",
    year = "1991"
}

@article{DELPHI:1991qnt,
    author = "Abreu, P. and others",
    collaboration = "DELPHI",
    title = "{Charged particle multiplicity distributions in restricted rapidity intervals in Z0 hadronic decays.}",
    doi = "10.1007/BF01560444",
    journal = "Z. Phys. C",
    volume = "52",
    pages = "271--281",
    year = "1991"
}

@article{H1:2020zpd,
    author = "Andreev, V. and others",
    collaboration = "H1",
    title = "{Measurement of charged particle multiplicity distributions in DIS at HERA and its implication to entanglement entropy of partons}",
    eprint = "2011.01812",
    archivePrefix = "arXiv",
    primaryClass = "hep-ex",
    reportNumber = "DESY-20-176",
    doi = "10.1140/epjc/s10052-021-08896-1",
    journal = "Eur. Phys. J. C",
    volume = "81",
    number = "3",
    pages = "212",
    year = "2021"
}

@article{UA5:1985kkp,
    author = "Alner, G. J. and others",
    collaboration = "UA5",
    title = "{A New Empirical Regularity for Multiplicity Distributions in Place of KNO Scaling}",
    reportNumber = "CERN-EP/85-62",
    doi = "10.1016/0370-2693(85)91492-3",
    journal = "Phys. Lett. B",
    volume = "160",
    pages = "199--206",
    year = "1985"
}

@article{UA5:1988gup,
    author = "Ansorge, R. E. and others",
    editor = "Kotthaus, R. and Kuhn, Johann H.",
    collaboration = "UA5",
    title = "{Charged Particle Multiplicity Distributions at 200-GeV and 900-GeV Center-Of-Mass Energy}",
    reportNumber = "CERN-EP-88-172",
    doi = "10.1007/BF01506531",
    journal = "Z. Phys. C",
    volume = "43",
    pages = "357",
    year = "1989"
}

@article{CMS:2010qvf,
    author = "Khachatryan, Vardan and others",
    collaboration = "CMS",
    title = "{Charged Particle Multiplicities in $pp$ Interactions at $\sqrt{s}=0.9$, 2.36, and 7 TeV}",
    eprint = "1011.5531",
    archivePrefix = "arXiv",
    primaryClass = "hep-ex",
    reportNumber = "CERN-PH-EP-2010-048, CMS-QCD-10-004",
    doi = "10.1007/JHEP01(2011)079",
    journal = "JHEP",
    volume = "01",
    pages = "079",
    year = "2011",
    number = "11"
}

@article{Grosse-Oetringhaus:2009eis,
    author = "Grosse-Oetringhaus, Jan Fiete and Reygers, Klaus",
    title = "{Charged-Particle Multiplicity in Proton-Proton Collisions}",
    eprint = "0912.0023",
    archivePrefix = "arXiv",
    primaryClass = "hep-ex",
    doi = "10.1088/0954-3899/37/8/083001",
    journal = "J. Phys. G",
    volume = "37",
    pages = "083001",
    year = "2010"
}

@article{ALICE:2017pcy,
    author = "Acharya, S. and others",
    collaboration = "ALICE",
    title = "{Charged-particle multiplicity distributions over a wide pseudorapidity range in proton-proton collisions at $\sqrt{s}=$ 0.9, 7, and 8 TeV}",
    eprint = "1708.01435",
    archivePrefix = "arXiv",
    primaryClass = "hep-ex",
    reportNumber = "CERN-EP-2017-192",
    doi = "10.1140/epjc/s10052-017-5412-6",
    journal = "Eur. Phys. J. C",
    volume = "77",
    number = "12",
    pages = "852",
    year = "2017"
}

@article{Konishi:1979cb,
    author = "Konishi, K. and Ukawa, A. and Veneziano, G.",
    title = "{Jet Calculus: A Simple Algorithm for Resolving QCD Jets}",
    reportNumber = "RL-79-026",
    doi = "10.1016/0550-3213(79)90053-1",
    journal = "Nucl. Phys. B",
    volume = "157",
    pages = "45--107",
    year = "1979"
}

@article{Bassetto:1979nt,
    author = "Bassetto, A. and Ciafaloni, M. and Marchesini, G.",
    title = "{Inelastic Distributions and Color Structure in Perturbative QCD}",
    reportNumber = "UTF-45",
    doi = "10.1016/0550-3213(80)90413-7",
    journal = "Nucl. Phys. B",
    volume = "163",
    pages = "477--518",
    year = "1980"
}

@article{Dokshitzer:1982ia,
    author = "Dokshitzer, Yuri L. and Fadin, Victor S. and Khoze, Valery A.",
    title = "{On the Sensitivity of the Inclusive Distributions in Parton Jets to the Coherence Effects in QCD Gluon Cascades}",
    reportNumber = "LENINGRAD-82-789",
    doi = "10.1007/BF01571703",
    journal = "Z. Phys. C",
    volume = "18",
    pages = "37",
    year = "1983"
}

@article{Bassetto:1987fq,
    author = "Bassetto, A.",
    title = "{KNO SCALING IN QCD JETS AND THE NEGATIVE BINOMIAL DISTRIBUTION}",
    reportNumber = "DFPD-9/87",
    doi = "10.1016/0550-3213(88)90426-9",
    journal = "Nucl. Phys. B",
    volume = "303",
    pages = "703--712",
    year = "1988"
}

@book{Dokshitzer:1991wu,
    author = "Dokshitzer, Yuri L. and Khoze, Valery A. and Mueller, Alfred H. and Troian, S. I.",
    title = "{Basics of perturbative QCD}",
    year = "1991",
    publisher = "Editions Frontieres",
    url = "https://www.lpthe.jussieu.fr/~yuri/BPQCD/cover.html"
}

@article{Dokshitzer:1993dc,
    author = "Dokshitzer, Yuri L.",
    title = "{Improved QCD treatment of the KNO phenomenon}",
    reportNumber = "LU-TP-93-3",
    doi = "10.1016/0370-2693(93)90121-W",
    journal = "Phys. Lett. B",
    volume = "305",
    pages = "295--301",
    year = "1993"
}

@article{Malaza:1984vv,
    author = "Malaza, E. D. and Webber, B. R.",
    title = "{QCD CORRECTIONS TO JET MULTIPLICITY MOMENTS}",
    reportNumber = "HEP 84/3",
    doi = "10.1016/0370-2693(84)90375-7",
    journal = "Phys. Lett. B",
    volume = "149",
    pages = "501--503",
    year = "1984"
}

@article{Dremin:2000ep,
    author = "Dremin, I. M. and Gary, J. W.",
    title = "{Hadron multiplicities}",
    eprint = "hep-ph/0004215",
    archivePrefix = "arXiv",
    reportNumber = "FIAN-TD31-00, UCRHEP-E273",
    doi = "10.1016/S0370-1573(00)00117-4",
    journal = "Phys. Rept.",
    volume = "349",
    pages = "301--393",
    year = "2001"
}

@article{Azzi:2019yne,
    author = "Azzi, P. and others",
    editor = "Dainese, Andrea and Mangano, Michelangelo and Meyer, Andreas B. and Nisati, Aleandro and Salam, Gavin and Vesterinen, Mika Anton",
    title = "{Report from Working Group 1}: {Standard Model Physics at the HL-LHC and HE-LHC}",
    eprint = "1902.04070",
    archivePrefix = "arXiv",
    primaryClass = "hep-ph",
    reportNumber = "CERN-LPCC-2018-03",
    doi = "10.23731/CYRM-2019-007.1",
    journal = "CERN Yellow Rep. Monogr.",
    volume = "7",
    pages = "1--220",
    year = "2019"
}

@book{Marzani:2019hun,
    author = "Marzani, Simone and Soyez, Gregory and Spannowsky, Michael",
    title = "{Looking inside jets: an introduction to jet substructure and boosted-object phenomenology}",
    eprint = "1901.10342",
    archivePrefix = "arXiv",
    primaryClass = "hep-ph",
    doi = "10.1007/978-3-030-15709-8",
    publisher = "Springer",
    volume = "958",
    year = "2019"
}

@article{Larkoski:2024uoc,
    author = "Larkoski, Andrew J.",
    title = "{QCD masterclass lectures on jet physics and machine learning}",
    eprint = "2407.04897",
    archivePrefix = "arXiv",
    primaryClass = "hep-ph",
    doi = "10.1140/epjc/s10052-024-13341-0",
    journal = "Eur. Phys. J. C",
    volume = "84",
    number = "10",
    pages = "1117",
    year = "2024"
}

@article{Vertesi:2020utz,
    author = "Vertesi, Robert and Gemes, Antal and Barnafoldi, Gergely Gabor",
    title = "{Koba-Nielsen-Olesen-like scaling within a jet in proton-proton collisions at LHC energies}",
    eprint = "2012.01132",
    archivePrefix = "arXiv",
    primaryClass = "hep-ph",
    doi = "10.1103/PhysRevD.103.L051503",
    journal = "Phys. Rev. D",
    volume = "103",
    number = "5",
    pages = "L051503",
    year = "2021"
}

@article{ATLAS:2011eid,
    author = "Aad, Georges and others",
    collaboration = "ATLAS",
    title = "{Properties of jets measured from tracks in proton-proton collisions at center-of-mass energy $\sqrt{s}=7$ TeV with the ATLAS detector}",
    eprint = "1107.3311",
    archivePrefix = "arXiv",
    primaryClass = "hep-ex",
    reportNumber = "CERN-PH-EP-2011-110",
    doi = "10.1103/PhysRevD.84.054001",
    journal = "Phys. Rev. D",
    volume = "84",
    pages = "054001",
    year = "2011"
}

@article{ATLAS:2019rqw,
    author = "Aad, Georges and others",
    collaboration = "ATLAS",
    title = "{Properties of jet fragmentation using charged particles measured with the ATLAS detector in $pp$ collisions at $\sqrt{s}=13$ TeV}",
    eprint = "1906.09254",
    archivePrefix = "arXiv",
    primaryClass = "hep-ex",
    reportNumber = "CERN-EP-2019-090",
    doi = "10.1103/PhysRevD.100.052011",
    journal = "Phys. Rev. D",
    volume = "100",
    number = "5",
    pages = "052011",
    year = "2019"
}

@article{Germano:2024ier,
    author = "Germano, G. R. and Navarra, F. S. and Wilk, G. and Wlodarczyk, Z.",
    title = "{Emergence of Koba-Nielsen-Olsen scaling in multiplicity distributions in jets produced at the LHC}",
    eprint = "2406.04856",
    archivePrefix = "arXiv",
    primaryClass = "hep-ph",
    doi = "10.1103/PhysRevD.110.034026",
    journal = "Phys. Rev. D",
    volume = "110",
    number = "3",
    pages = "034026",
    year = "2024"
}

@article{Dokshitzer:1987nm,
    author = "Dokshitzer, Yuri L. and Khoze, Valery A. and Troian, S. I. and Mueller, Alfred H.",
    title = "{QCD Coherence in High-Energy Reactions}",
    reportNumber = "CU-TP-374",
    doi = "10.1103/RevModPhys.60.373",
    journal = "Rev. Mod. Phys.",
    volume = "60",
    pages = "373",
    year = "1988"
}

@article{Larkoski:2013eya,
    author = "Larkoski, Andrew J. and Salam, Gavin P. and Thaler, Jesse",
    title = "{Energy Correlation Functions for Jet Substructure}",
    eprint = "1305.0007",
    archivePrefix = "arXiv",
    primaryClass = "hep-ph",
    reportNumber = "MIT-CTP-4446, CERN-PH-TH-2013-066, LPN13-026",
    doi = "10.1007/JHEP06(2013)108",
    journal = "JHEP",
    volume = "06",
    pages = "108",
    year = "2013",
    number = "13"
}

@article{Buckley:2014ana,
    author = {Buckley, Andy and Ferrando, James and Lloyd, Stephen and Nordstr\"om, Karl and Page, Ben and R\"ufenacht, Martin and Sch\"onherr, Marek and Watt, Graeme},
    title = "{LHAPDF6: parton density access in the LHC precision era}",
    eprint = "1412.7420",
    archivePrefix = "arXiv",
    primaryClass = "hep-ph",
    reportNumber = "GLAS-PPE-2014-05, MCNET-14-29, IPPP-14-111, DCPT-14-222",
    doi = "10.1140/epjc/s10052-015-3318-8",
    journal = "Eur. Phys. J. C",
    volume = "75",
    pages = "132",
    year = "2015"
}

@article{Liu:2022bru,
    author = "Liu, Yizhuang and Nowak, Maciej A. and Zahed, Ismail",
    title = "{Mueller\textquoteright{}s dipole wave function in QCD: Emergent Koba-Nielsen-Olesen scaling in the double logarithm limit}",
    eprint = "2211.05169",
    archivePrefix = "arXiv",
    primaryClass = "hep-ph",
    doi = "10.1103/PhysRevD.108.034017",
    journal = "Phys. Rev. D",
    volume = "108",
    number = "3",
    pages = "034017",
    year = "2023"
}

@article{Liu:2023eve,
    author = "Liu, Yizhuang and Nowak, Maciej A. and Zahed, Ismail",
    title = "{Universality of Koba-Nielsen-Olesen scaling in QCD at high energy and entanglement}",
    eprint = "2302.01380",
    archivePrefix = "arXiv",
    primaryClass = "hep-ph",
    month = "2",
    year = "2023",
    journal = "-",
}

@article{Bierlich:2022pfr,
    author = "Bierlich, Christian and others",
    title = "{A comprehensive guide to the physics and usage of PYTHIA 8.3}",
    eprint = "2203.11601",
    archivePrefix = "arXiv",
    primaryClass = "hep-ph",
    reportNumber = "LU-TP 22-16, MCNET-22-04, FERMILAB-PUB-22-227-SCD",
    doi = "10.21468/SciPostPhysCodeb.8",
    journal = "SciPost Phys. Codeb.",
    volume = "2022",
    pages = "8",
    year = "2022"
}

@article{Cacciari:2008gp,
    author = "Cacciari, Matteo and Salam, Gavin P. and Soyez, Gregory",
    title = "{The anti-$k_t$ jet clustering algorithm}",
    eprint = "0802.1189",
    archivePrefix = "arXiv",
    primaryClass = "hep-ph",
    reportNumber = "LPTHE-07-03",
    doi = "10.1088/1126-6708/2008/04/063",
    journal = "JHEP",
    volume = "04",
    pages = "063",
    year = "2008",
    number = "08"
}

@article{Cacciari:2011ma,
    author = "Cacciari, Matteo and Salam, Gavin P. and Soyez, Gregory",
    title = "{FastJet User Manual}",
    eprint = "1111.6097",
    archivePrefix = "arXiv",
    primaryClass = "hep-ph",
    reportNumber = "CERN-PH-TH-2011-297",
    doi = "10.1140/epjc/s10052-012-1896-2",
    journal = "Eur. Phys. J. C",
    volume = "72",
    pages = "1896",
    year = "2012"
}

@article{Azimov:1984np,
    author = "Azimov, Yakov I. and Dokshitzer, Yuri L. and Khoze, Valery A. and Troyan, S. I.",
    title = "{Similarity of Parton and Hadron Spectra in QCD Jets}",
    reportNumber = "LENINGRAD-84-942",
    doi = "10.1007/BF01642482",
    journal = "Z. Phys. C",
    volume = "27",
    pages = "65--72",
    year = "1985"
}

@article{Kang:2023zdx,
    author = "Kang, Zhong-Bo and Kao, Robert and Larkoski, Andrew J.",
    title = "{Multiplicity scaling of fragmentation function}",
    eprint = "2305.13359",
    archivePrefix = "arXiv",
    primaryClass = "hep-ph",
    doi = "10.1103/PhysRevD.109.054039",
    journal = "Phys. Rev. D",
    volume = "109",
    number = "5",
    pages = "054039",
    year = "2024"
}

@article{Medves:2022ccw,
    author = "Medves, Rok and Soto-Ontoso, Alba and Soyez, Gregory",
    title = "{Lund and Cambridge multiplicities for precision physics}",
    eprint = "2205.02861",
    archivePrefix = "arXiv",
    primaryClass = "hep-ph",
    reportNumber = "OUTP-22-07P",
    doi = "10.1007/JHEP10(2022)156",
    journal = "JHEP",
    volume = "10",
    pages = "156",
    year = "2022",
    number = "22"
}

@article{Medves:2022uii,
    author = "Medves, Rok and Soto-Ontoso, Alba and Soyez, Gregory",
    title = "{Lund multiplicity in QCD jets}",
    eprint = "2212.05076",
    archivePrefix = "arXiv",
    primaryClass = "hep-ph",
    reportNumber = "CERN-TH-2022-205, OUTP-22-13P",
    doi = "10.1007/JHEP04(2023)104",
    journal = "JHEP",
    volume = "04",
    pages = "104",
    year = "2023",
    number = "23"
}

@article{DiasdeDeus:1997di,
    author = "Dias de Deus, J. and Pajares, C. and Salgado, C. A.",
    title = "{Moment analysis, multiplicity distributions and correlations in high-energy processes: Nucleus-nucleus collisions}",
    eprint = "hep-ph/9702398",
    archivePrefix = "arXiv",
    reportNumber = "US-FT-6-97",
    doi = "10.1016/S0370-2693(97)00677-1",
    journal = "Phys. Lett. B",
    volume = "407",
    pages = "335--340",
    year = "1997"
}

@article{Armesto:2008qe,
    author = "Armesto, Nestor and Pajares, Carlos and Quiroga-Arias, Paloma",
    editor = "Armesto, Nestor and Pajares, Carlos and Salgado, Carlos A. and Wiedemann, Urs A.",
    title = "{Medium dependence of multiplicity distributions in MLLA}",
    eprint = "0809.4428",
    archivePrefix = "arXiv",
    primaryClass = "hep-ph",
    doi = "10.1140/epjc/s10052-008-0829-6",
    journal = "Eur. Phys. J. C",
    volume = "61",
    pages = "779--784",
    year = "2009"
}

@article{Armesto:2009aqm,
    author = "Armesto, Nestor and Pajares, Carlos and Quiroga-Arias, Paloma",
    title = "{Jet multiplicity distributions: Medium dependence in MLLA}",
    doi = "10.1140/epjc/s10052-009-0899-0",
    journal = "Eur. Phys. J. C",
    volume = "62",
    pages = "151--157",
    year = "2009"
}

@article{Ma:2003vs,
    author = "Ma, G. L. and others",
    title = "{Delta scaling and information entropy in ultrarelativistic nucleus nucleus collisions}",
    eprint = "nucl-th/0306030",
    archivePrefix = "arXiv",
    doi = "10.1088/0256-307X/20/7/312",
    journal = "Chin. Phys. Lett.",
    volume = "20",
    pages = "1013--1016",
    year = "2003"
}

@article{Ma:2004eqa,
    author = "Ma, Y. G. and others",
    editor = "Cleymans, J. and Vilakazi, Z. and Steinberg, P.",
    title = "{Delta-scaling and heat capacity in relativistic ion collisions}",
    eprint = "nucl-th/0411112",
    archivePrefix = "arXiv",
    doi = "10.1088/0954-3899/31/6/082",
    journal = "J. Phys. G",
    volume = "31",
    pages = "S1179--S1182",
    year = "2005"
}

@article{Escobedo:2016vba,
    author = "Escobedo, Miguel A. and Iancu, Edmond",
    title = "{Multi-particle correlations and KNO scaling in the medium-induced jet evolution}",
    eprint = "1609.06104",
    archivePrefix = "arXiv",
    primaryClass = "hep-ph",
    doi = "10.1007/JHEP12(2016)104",
    journal = "JHEP",
    volume = "12",
    pages = "104",
    year = "2016",
    number = "16"
}

@unpublished{DCMSW2025,
    author = "Duan, Xiang-Pan and Chen, Lin and Ma, Guo-Liang and Salgado, Carlos A. and Wu, Bin",
    title = "{Multiplicity distributions in QCD jets and jet topics}",
    eprint = "2509.06158",
    archivePrefix = "arXiv",
    primaryClass = "hep-ph",
    month = "9",
    year = "2025"
}

@article{ALEPH:1995qic,
    author = "Buskulic, D. and others",
    collaboration = "ALEPH",
    title = "{Measurements of the charged particle multiplicity distribution in restricted rapidity intervals}",
    reportNumber = "CERN-PPE-95-082, CERN-PPE-95-82",
    doi = "10.1007/BF02907382",
    journal = "Z. Phys. C",
    volume = "69",
    pages = "15--26",
    year = "1995"
}

@article{Giovannini:1997ce,
    author = "Giovannini, Alberto and Ugoccioni, Roberto",
    editor = "Capon, G. and Khoze, Valery A. and Pancheri, G. and Sansoni, A.",
    title = "{Soft and semihard components structure in multiparticle production in high-energy collisions}",
    eprint = "hep-ph/9710361",
    archivePrefix = "arXiv",
    reportNumber = "DFTT-65-97",
    doi = "10.1016/S0920-5632(98)00343-0",
    journal = "Nucl. Phys. B Proc. Suppl.",
    volume = "71",
    pages = "201--210",
    year = "1999"
}

@article{Metodiev:2018ftz,
    author = "Metodiev, Eric M. and Thaler, Jesse",
    title = "{Jet Topics: Disentangling Quarks and Gluons at Colliders}",
    eprint = "1802.00008",
    archivePrefix = "arXiv",
    primaryClass = "hep-ph",
    reportNumber = "MIT-CTP-4979",
    doi = "10.1103/PhysRevLett.120.241602",
    journal = "Phys. Rev. Lett.",
    volume = "120",
    number = "24",
    pages = "241602",
    year = "2018"
}

@article{Komiske:2018vkc,
    author = "Komiske, Patrick T. and Metodiev, Eric M. and Thaler, Jesse",
    title = "{An operational definition of quark and gluon jets}",
    eprint = "1809.01140",
    archivePrefix = "arXiv",
    primaryClass = "hep-ph",
    reportNumber = "MIT-CTP 5042",
    doi = "10.1007/JHEP11(2018)059",
    journal = "JHEP",
    volume = "11",
    pages = "059",
    year = "2018"
}
\end{document}